\documentclass{article}

\usepackage{PRIMEarxiv}

\usepackage[utf8]{inputenc} 
\usepackage[T1]{fontenc}    
\usepackage{hyperref}       
\usepackage{url}            
\usepackage{booktabs}       
\usepackage{amsfonts}       
\usepackage{nicefrac}       
\usepackage{microtype}      
\usepackage{lipsum}
\usepackage{fancyhdr}       
\usepackage{graphicx}       
\usepackage{xcolor}
\usepackage{algorithm}
\usepackage{algpseudocode}
\usepackage{subcaption}
\usepackage{booktabs}
\usepackage{graphicx}
\usepackage{array}
\usepackage{enumitem}
\usepackage{calc} 
\graphicspath{{media/}}     

\title{Evidence-centered Assessment for Writing with Generative AI
}

\author{
  Yixin Cheng \\
  Monash University\\
  \texttt{yixin.cheng@monash.edu} \\
   \And
  Kayley Lyons \\
  University of Melbourne\\
  \texttt{kayley.lyons@unimelb.edu.au} \\
  \AND
  Guanliang Chen \\
  Monash University\\
  \texttt{guanliang.chen@monash.edu} \\
  \And
  Dragan Ga\v{s}evi\'c \\
  Monash University\\
  \texttt{dragan.gasevic@monash.edu} \\
  \And
  Zachari Swiecki \\
  Monash University\\
  \texttt{zach.swiecki@monash.edu} \\
}

\begin{document}
\maketitle

\begin{abstract}
We propose a learning analytics-based methodology for assessing the collaborative writing of humans and generative artificial intelligence. Framed by the evidence-centered design, we used elements of knowledge-telling, knowledge transformation, and cognitive presence to identify assessment claims; we used data collected from the \textit{CoAuthor} writing tool as potential evidence for these claims; and we used epistemic network analysis to make inferences from the data about the claims. Our findings revealed significant differences in the writing processes of different groups of \textit{CoAuthor} users, suggesting that our method is a plausible approach to assessing human-AI collaborative writing.
\end{abstract}

\keywords{Generative Artificial Intelligence \and Assessment \and Evidence-centered Design \and Epistemic Network Analysis}

\section{Introduction}
Effectively communicating ideas via writing is a critical human skill. Every day, many of us send text messages, draft emails, and make notes; in many specialised domains, such as research, writing is a core form of discourse. The process of writing has, of course, changed over time; writing tools have transformed from mere storage receptacles to tools that help us craft more effective writing, such as word processors, spellcheckers, and grammar checkers \cite{naber2003}. Recently, however, there has been a step-change in tools for writing. Whereas prior tools helped us to save and process our own writing, tools-based on \emph{generative artificial intelligence} (GAI) can now compose writing for us.

This technological advance has already had far reaching implications for society. Arguably, these implications have been---and will continue to be---the most profound for education. Education relies, in part, on \textit{assessment}. Broadly speaking, we want students to learn skills that will be valuable to their well-being and the well-being of society. But to know and communicate whether learning has occurred, we need to be able to \emph{evidence} that learning. Assessments are the structures that make such inferences about learning possible \cite{Mislevy2003}. A key feature of assessments is that they assign credit to both pragmatic actions---actions that advance toward a goal state---and epistemic actions---actions that reduce cognitive effort \cite{Clark1998}. However, when students use tools like \textit{ChatGPT} \cite{openai2023gpt4} to help them write, these actions are mixed between the human and the tool \cite{wertsch_mind_1998}, making the assignment of credit difficult. How, for example, should a teacher assess writing crafted partly by a human and partly by AI?

In the last year, educational institutions across the world have scrambled to put policies in place that seek to address questions like the one above. For example, the Tertiary Education Quality and Standards Agency (TEQSA), responsible for regulating and the quality of all providers of higher education in Australia, recently convened a panel of experts to produce a whitepaper on “assessment reform in the age of artificial intelligence” that outlines broad changes institutes of higher education will be expected to make in response to AI \cite{Lodge2023}. Rather than cut students off from AI, its authors argue that assessments should prepare students for a world where collaboration with AI is commonplace by redesigning assessments to emphasize: (a) appropriate and authentic engagement with AI; (b) the process of learning; and (c) opportunities for students to work with each other and AI.

We agree with these recommendations. In turn, we argue that there is a need to re-develop assessments to account for interactions between humans and AI in ways that afford reasoned arguments about learning claims from verifiable and persistent evidence. In this paper, we propose and test such an assessment informed by (a) the \textit{evidence-centered design} (ECD) assessment framework (b) extant theories of the cognitive processes involved in writing (c) recent interfaces to GAI, and (d) learning analytic process models. Our results suggest that this method can find expected differences between the processes people use when writing with GAI, such as differences between genres. This work is a proof of concept that contributes to a deeper understanding of human-AI collaboration and suggests a path forward for innovative assessments in this new age of AI.

\section{Background and Theory}

Educational assessments are a means for determining whether and how students have learned. While several frameworks for assessment design have been proposed, the ECD framework has proven to be particularly successful and adaptable to a variety of domains \cite{Mislevy2003}. ECD defines a Conceptual Assessment Framework that incorporates three high-level models: a \emph{student model}, \emph{task model}, and \emph{evidence model}. Together, these components suggest a coherent and logical approach for designing assessments that are aligned with the constructs they intend to measure.

\subsection{ECD Student Model}
\label{subsec:student model}
The student model identifies and describes the claims about learning that the assessment aims to measure. Our study aimed to assess the \textit{processes} students enact while writing with GAI as opposed to the quality of the final writing \textit{product}. There is a long history of assessing writing ability---as well as learning more generally---in terms of products like essays, posts, and articles. For example, automated essay scoring \cite{Ke2019} and writing analytics \cite{Shibani2019} have developed into active sub-fields of psychometrics and learning analytics, respectively. Such research traditionally uses natural language processing (NLP) techniques to assess writing quality and rhetorical structure. More recently, work has transitioned from machine learning and traditional NLP models to analyses based on generative language models \cite{Wang2022,Iqbal2023}.

While assessing products is undoubtedly important, assessing process has been recognised as valuable in the learning sciences and learning analytics communities because it offers a more complete view of student abilities, facilitates effective feedback, and aids in creating adaptive scaffolds for personalized needs \cite{Gasevic2015,Li2023}. We argue that a process-oriented view of assessment will help to resolve the many of the challenges current educators face when incorporating or accounting for GAI in their assessment design. Much of the worry surrounding assessment and GAI is centered on the only evidence of learning being a product with unknown providence \cite{Swiecki2022}. Altering assessment to instead focus on writing processes will make it clearer who (the student or the AI) produced which actions and how these actions might positively or negatively impact student learning.

While few studies of how people co-write with GAI have been reported to date, the interfaces of various GAI tools suggest particular kinds of interaction patterns. \textit{ChatGPT} \cite{openai2023gpt4}, for example, includes a text area for prompting the AI and the ability to copy text, regenerate text, and like or dislike a response. Similarly, Google's \emph{Bard} \cite{bard2023} lets users modify responses within the interface and export them to emails or word documents directly from the tool. These affordances suggest that when using GAI writers may engage in varied processes of prompting, as well as exploring, using, and modifying AI generated responses.

\subsubsection{Knowledge Telling and Knowledge Transforming in Writing}
Bereiter and Scardamalia's work on knowledge building offers a lens with which to view these kinds of interactive writing behaviors \cite{Bereiter1987}. They distinguish between \emph{knowledge telling}---a linear process of writing where individuals focus on transcribing information without deep engagement or critical restructuring---and \textit{knowledge transformation}, a more critical engagement that restructures arguments thoughtfully, synthesizes different viewpoints, and creates narratives that are both cohesive and compelling. Raković \cite{mladen2019thesis} extended this work to argue that knowledge transformation is a set of actions whereby writers comprehend, evaluate, and select information sources to fostering novel connections among previously disconnected fragments of knowledge to facilitate learning.

In the context of writing with GAI, we operationalize knowledge telling and knowledge transformation in terms of how learners select, modify, and apply AI-generated text. More specifically, knowledge telling might appear as learners accepting suggestions from the AI without alteration, following the path laid out by the AI at any given moment. Knowledge transformation, on the other hand, can be seen when users take AI suggestions as a basis for revision and modify them to fit their needs. Learners may also engage in actions such as declining initial GAI suggestions and seeking alternatives several times; or they might deliberate between different sets of suggestions. Such example defy a simple categorization under knowledge telling or knowledge transformation. However, we argue below that the \textit{cognitive presence} framework can help us to understand these interactions.

\subsubsection{Cognitive Presence}
Garrison situated \cite{garrison1999} cognitive presence within the Community of Inquiry (CoI) framework---a theoretical framework that promotes educational experience through computer-mediated communication. Cognitive presence refers to the extent to which learners are able to construct and confirm meaning through communication in a specific sociocultural context. It is manifested through a four-phase cycle that includes a \emph{triggering event}---the initial encounter with a problem or question that ignites curiosity; \emph{exploration}---where learners actively seek information to deepen their understanding to the question; \emph{integration}---where gathered information is synthesized to formulate coherent responses; and finally, \emph{resolution}---where synthesized knowledge is used to propose a solution \cite{garrison1999}. Prior work has largely focused on the context of online discussion---for example, automatically identifying cognitive presence in messages \cite{Hu2021} and associating cognitive presence with particular speech acts \cite{Iqbal2022}. However, cognitive presence has yet to be investigated in the context of collaborative writing between humans and AI.

In this study, we operationalize cognitive presence in terms of interactions that take place while co-writing with GAI. Triggering events may occur when learners approach GAI with an initial question or problem they encounter while they are writing. Exploration may occur as learners actively engage with the GAI to brainstorm or search through various suggestions. Integration may occur as learners synthesize the ideas and suggestions received from the AI, working towards constructing a cohesive piece of writing. In this way, integration is similar to knowledge transformation as described above. Finally, resolution may occur when learners fine-tune their piece with the aid of the AI, seeking advice on polishing the language and structure to reach a satisfactory final product.

The combination of knowledge telling, knowledge transformation, and cognitive presence suggests a set of claims to include in the student model for assessing the processes involved in co-writing with GAI: using generative AI responses as is (knowledge telling); modifying generative AI responses to fit your own goals (knowledge transformation/integration); requesting information from GAI (triggering); comparing different GAI outputs (exploration); and arriving at a final product that aligns with the human writer's intent (resolution).

\subsection{ECD Task model}
The task model specifies the tasks and task environment students will interact with to evidence their learning. As mentioned before, several commercial interfaces to GAI exist. These tools allow individuals to prompt GAI and integrate the outputs into their own writing. While powerful, they fall short of requirements of an efficient task model because the data they capture is not easily accessible for assessment purposes. This lack of the accessibility leaves educators with few options: either attempt to ban students from using GAI---which is likely to fail because it is notoriously difficult to invigilate students' online activity \cite{Cramp2019}---or try to document students' use of GAI---which is difficult and time-consuming. It is hard to imagine a scenario where a teacher reviews hours of videos or analyzes numerous screenshots of AI interactions for each of their students. A more viable solution would be platforms designed to automatically capture and preserve evidence related to writing with GAI.

One such platform is \textit{CoAuthor} \cite{Lee2022}---an online tool designed to (a) afford interactions with GAI and (b) passively capture those interactions for analysis. \textit{CoAuthor} was designed as a test-bed for investigating human-AI collaborative writing. It provides users with writing prompts either for argumentative writing (e.g. essays) or creative writing (e.g. stories) and asks users to compose a response. As they write, they can seek suggestions from GAI that continue their writing. Users are free to explore, accept, or dismiss suggestions and modify them if accepted. While users are constrained in their interactions with GAI compared to commercial tools---for example, they cannot prompt the AI with their own questions---\textit{CoAuthor} provides a suitable task environment for assessment purposes because it automatically logs and makes accessible fine-grained interactions with the system. All keystrokes, cursor movements, and interactions with GAI are logged in both a human and machine readable format. Here, we use existing data collected as part of the \textit{CoAuthor} project to demonstrate what a valid assessment for writing with GAI might look like.

\subsection{ECD Evidence Model}
The ECD evidence model describes how students' responses to the task will be related to the claims in the student model. Recently, the sub-field of \textit{writing analytics} has developed a variety of methods for evidencing claims from writing data \cite{Gibson2022}. To understand writing processes, trace data (e.g. keystroke logging, mouse movement, eye tracking, etc.) \cite{Swiecki2022} are typically analyzed using \textit{process models}. For example, recent research \cite{Guo2019,zhang2021,Mladen2023-jcal} has explored students' cognitive and metacognitive processes during essay writing via trace data by applying techniques like semi-Markov processes \cite{Limnios2001}.

To better understand writing processes and process models, researchers often use \textit{process graphs}, which show the relationships or connections between different writing behaviors. For example, Leijten and Waes \cite{Leijten2013} used the process graph \textit{Inputlog}\footnote{\url{https://www.inputlog.net}} to visualize writing progression over time. They further adapted work into a network visualization with \textit{Pajek}\footnote{\url{http://mrvar.fdv.uni-lj.si/pajek/}} to highlight the key information sources used during writing. Other researchers have used representations called \textit{revision maps} to investigate writing processes. For example, Southavilay et al. \cite{Southavilay2013} developed a revision map to visually track changes in paragraphs of a collaborative document over time using color-coded rectangles to signify the nature and extent of edits. Likewise, Shibani and colleagues \cite{Shibani2018} created a revision graph to analyze the editing process in essay drafts using nodes to represent sentences and edges to depict the organizational changes across drafts. 

Building on revision maps and process graphs, Shibani et al. \cite{shibani2023} introduced \textit{CoAuthorViz}. This more advanced process graph visualizes co-writing behaviors with GPT-3 suggestions, drawing upon trace data from \emph{CoAuthor}. Different from traditional process graphs, it emphasizes three key constructs: "Suggestion Selection", "Suggestion Modification", "Empty GPT-3 Call", and "User Text Addition" detailing the sequence of writing actions at the sentence level. These graphical representations offer insights into sentence-level differences in compositions and revisions.

The process models and graph approaches described above are commendable for their focus on the relationships between features that characterise writing; however, they have important limitations as components of an evidence model. First, traditional process graphs and revision maps are useful for understanding individual or aggregated writing processes; however they do a poor job of effectively \textit{comparing} processes between individuals or groups. This is primarily due to their lack of a coordinated semantic space for network comparisons. Second, while Markov-models, traditional process graphs, and revision maps excel at showing detailed relationships, they lack corresponding \textit{summary statistics} that can be used to quantify the network interpretation. Such summary statistics are useful because they afford process comparisons at scale using standard statistical techniques.

The process graph introduced in \textit{CoAuthorViz} distinguishes itself by analyzing writing across two groups of users and incorporating summary statistics. Despite these advancements, the tool has several notable limitations. First, the constructs included in the model are relatively broad and may therefore miss the fine-grained behaviors essential for analyzing writing in a nuanced and theoretically driven way. For example, "Suggestion Selection" consists of three distinct behaviors---suggestion seeking, exploration, and acquirement, but these are aggregated together and analyzed as a single construct that is not explicitly related to theories that describe the cognitive aspects of writing. Second, their comparison of  group of users was based on the final written product, not writing processes.

To address these limitations, we used epistemic network analysis (ENA) as the major component of our evidence model. ENA is a widely used learning analytic technique for modeling actions via network models \cite{Shafferruis2017}. ENA coordinates multiple networks in the same embedding space to facilitate visual and statistical comparisons of learning processes at scale. While ENA and related techniques have been used to model metacognitive processes during writing tasks \cite{fan2023}, no work has explored ENA as method for evidencing the cognitive behaviors inherent to co-writing with GAI.

\subsection{Research Questions}
Our study aimed to demonstrate an assessment method for human-AI collaborative writing. Framed by ECD, we used elements of knowledge-telling, knowledge transformation, and cognitive presence to identify claims for our student model (see hypotheses below); we used data collected from the \textit{CoAuthor} writing tool as a proxy for our task model; and we used ENA to evidence claims for the student model. Ideally, this assessment approach would distinguish high-quality human-AI collaborative writing processes from low-quality ones. Unfortunately, the \textit{CoAuthor} dataset does not include writing quality scores that could be correlated with writing process measures for this purpose. However, the dataset does contain a number of conditions that should be distinguishable with a valid assessment method. In particular, written products and trace data were divided by how much ownership---i.e., the proportion of content generated by the human author versus the GAI---the human had over the final product; whether the author was responding to a creative writing prompt or an argumentative writing prompt; and whether the GAI had a high or low \textit{temperature} setting---that is, whether its outputs had high or low textual variability. To provide a proof of concept for our assessment method, we compared these conditions according to the following research questions:

\begin{description}[leftmargin=32pt, labelindent=10pt, style=multiline]
    \item[RQ1] Did authors with higher ownership over the final product write differently than those with lower ownership over the final product?
    \item[RQ2] Did authors who responded to creative writing prompts write differently than those who responded to argumentative prompts?
    \item[RQ3] Did authors who use GAI with a higher temperature setting—that is, more variable output—write differently than those who used a lower temperature setting?
\end{description}

Regarding RQ1, we hypothesized that: (1) authors who had lower ownership over the final product would have writing processes more characterised by knowledge telling and triggering events because they sought and relied more on ready-made suggestions provided by the GAI; (2) conversely, authors with higher ownership over the final product would have processes more characterised by knowledge transformation, exploration, integration, and resolution because they tended to adapt GAI suggestions to better fit their writing.

To address RQ2, we built upon genre theory \cite{chandler1997}, which suggests major differences between fiction (e.g. creative) and non-fiction (e.g. argumentative) writing. Namely, fiction writing is inherently less bound to a "reality status" or tight relationship to the real world. Therefore, authors in this genre are more free to explore the boundaries of imagination and creativity without the limits of factual accuracy. On the other hand, non-fiction writing is grounded in reality, requiring a structured approach to presenting logical arguments based on facts and evidence. Thus, for RQ2 we hypothesized that: (3) authors responding to creative writing prompts would have writing processes more characterised by knowledge telling, triggering events, and exploration as they would likely need to adapt GAI responses less to maintain a reality status; and (4) authors responding to argumentative writing prompts would have writing processes more characterised by knowledge transformation, integration, and resolution as they may have needed to align GAI outputs (which are known to include falsehoods \cite{Ji2023}) with reality. 

High temperature settings mean that GAI will tend to produce varied outputs to the same request, while low temperature settings mean that GAI will tend to produce repetitive outputs. Regarding RQ3, we hypothesized that: (5) authors interacting with lower temperature GAI would have processes more characterized by knowledge transformation, integration, and resolution as they would need to actively think and rework the available suggestions; (6) while those interacting with higher temperature GAI would have processes more characterized by exploration and knowledge telling as they might find it easier to select suggestions that align closely with their intended message without significantly transforming the information.

\section{Method}
\begin{figure*}[h]
  \centering
  \includegraphics[width=\linewidth]{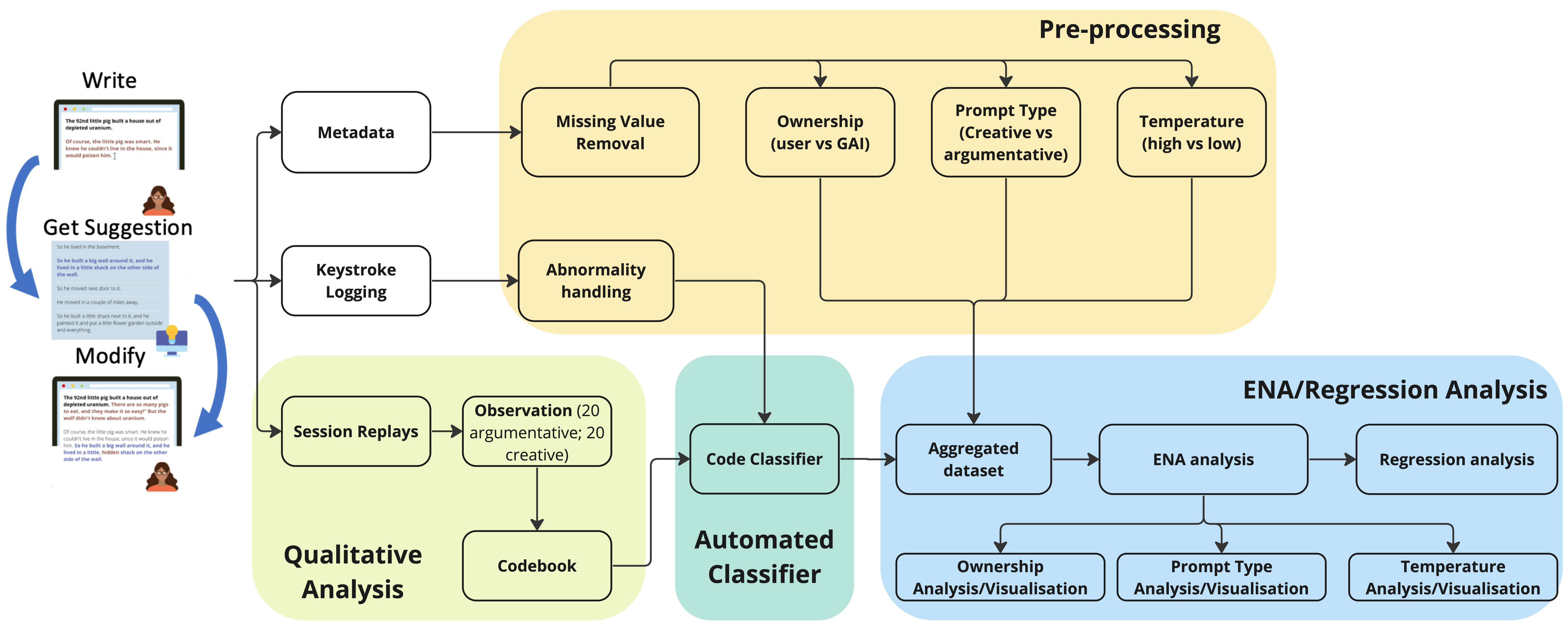}
  \caption{Methodology overview (the leftmost image sourced from \cite{Lee2022}) }
  \label{method-overall}
\end{figure*}

As shown in Figure \ref{method-overall}, our methodological steps included (a) pre-processing the keystroke-level data from \textit{CoAuthor}; (b) qualitatively analyzing these data to derive codes related to knowledge telling, transformation, cognitive presence, and general writing behaviors for subsequent analysis; (c) the development, validation, and application of classifiers for automatic assignment of the codes to the data; (d) applying ENA to the data to create networks that described the writing processes of authors in different experimental conditions; and (e) using summary statistics from ENA to statistically compare the experimental conditions via mixed-effects regression analysis.

\subsection{Data}
We used data collected from the \textit{CoAuthor} project as a proxy for the task model of our proposed assessment methodology. These data include digital records of 1,445 writing sessions from 60 authors recruited from Amazon Mechanical Turk. Authors were asked to respond to one or more prompts. Prompts could be from the creative writing condition, which were sampled from the r/WritingPrompts subreddit, or from the argumentative writing condition, which were sampled from a \textit{New York Times} prompt library. During the writing sessions, authors interacted with a user interface that included a standard text editor as well as a tab that allowed them to request up to five suggestions from GPT-3 for continuing the currently written text. 

Overall, elements of the dataset we used included two distinct types of data:

\begin{itemize}
    \item{\textbf{Trace Data}}: Sequential actions and events throughout the writing session. These included user actions such as text insertion, deletion, and cursor movement, as well as system interactions like opening, getting, and closing suggestions. Furthermore, it traced the progression of the written content on an event-by-event basis. 
    \item{\textbf{Metadata}}: Details such as the session time, participant ID, session ID, prompt condition, prompt ID, GPT temperature, as well as metrics that summarise the interactions between the user and GPT such as ownership---that is the percentage of the final written text produced by the human author vs. GPT-3. 
\end{itemize}

\subsection{Pre-processing}
The original data included 830 creative writing sessions and 615 argumentative writing sessions which averaged 1,862 events per session. We removed eight sessions from the analysis due to participants removing the original prompt and commencing a different writing piece, and one session due to missing metadata about identification of the author. In addition, we identified and manually corrected or excluded events with formatting errors. Our final dataset consisted of 822 creative writing sessions and 614 argumentative writing sessions that included more than 2.6M events. For sessions using argumentative writing prompts, low/high temperature values were 0.2 and 0.9; for creative writing prompts they were 0.3 and 0.75. Author and GAI ownership was defined using a median split on the percentage of characters authored by a human (mdn = 76\%). 

\subsection{Qualitative Analysis}
ENA requires a coded dataset for analysis---that is, data where the relevant actions have been labelled. To derive codes, we combined a top-down approach informed by theory with a bottom-up approach that sought to identify relevant themes in the data without explicit reference to theoretical frameworks\cite{Shaffer2021}. We qualitatively investigated 40 randomly sampled sessions for code identification. Two authors discussed and refined these codes to arrive at the final list below in Table \ref{tab:codebook}.

\begin{table*}[h]
\small
\centering
\caption{Qualitative codes, definitions, and identifiers.}
\resizebox{\textwidth}{!}{\begin{tabular}{p{3cm}p{5cm}p{6cm}}
\hline
\textbf{Code} & \textbf{Definition} & \textbf{Identifiers} \\ \hline
\textsc{compose} & Writing new content at the end of the existing text & event name is "text-insert" in the end of text (space removed where applicable) \\ \hline
\textsc{relocate} & Rearranging sentences & The index of same sentence changes in last-current document\\ \hline
\textsc{reflect} & Revising the content after or near completing a draft & Revise content after finishing 90\% content \\ \hline
\textsc{seek suggestion}& Obtaining suggestions from GPT & Event name is 'suggestion-get' \\ \hline
\textsc{dismiss suggestion} & Dismissing suggestions from GPT & Event name is 'suggestion-close' and event source is 'user' \\ \hline
\textsc{accept suggestion} & Accepting a suggestion from GPT & Event name is 'suggestion-select' \\ \hline
\textsc{hover suggestion} & Hovering over the suggestions & Event name is 'suggestion-hover' \\ \hline
\textsc{cursor forward} & Moving the position of cursor forward & Event name is "cursor-forward" \\ \hline
\textsc{cursor backward} & Moving the position of cursor backward & Event name is "cursor-backward"\\ \hline
\textsc{cursor select} & Selecting text & Event name is "cursor-select"  \\ \hline
\textsc{revise user} & Revising content they wrote & The inserts or deletes in-text content and ownership of revising sentence is 'user' \\ \hline
\textsc{revise suggestion} & Revising a GPT suggestion & The user inserts or deletes in-text content and ownership of revising sentence is 'api' \\ \hline
\textsc{low modification} & Making minor adjustments to sentences without altering their core meaning & Sentence semantic similarity > 0.8 \\ \hline
\textsc{high modification} & Making significant changes to sentence meaning & Sentence semantic similarity < 0.8 \\ \hline
\end{tabular}}
\label{tab:codebook}
\end{table*}

In our analysis, codes were related to the student model in terms of the presence or absence of co-occurrences or \emph{connections} between them (see \ref{sub:ENA}). We interpreted \emph{knowledge telling} in terms of connections among \textsc{seek suggestion}, \textsc{accept suggestion}, \textsc{revise suggestion}, and \textsc{low modification}; \emph{knowledge transformation} and \emph{integration} in terms of connections among \textsc{seek suggestion}, \textsc{accept suggestion}, \textsc{revise suggestion}, \textsc{high modification}, and \textsc{relocate}; \emph{triggering events} in terms of connections to \textsc{seek suggestion} and \textsc{accept suggestion}; \emph{exploration} in terms of connections among \textsc{seek suggestion}, \textsc{accept suggestion}, \textsc{hover suggestion}, and \textsc{dismiss suggestion}; and \emph{resolution} in terms of connections among \textsc{seek suggestion}, \textsc{accept suggestion}, and \textsc{reflect}. In addition to codes that we mapped to knowledge telling/transformation/cognitive presence, our qualitative analysis suggested that including codes related to composing, revising one's own text, and cursor movements would be useful for more completely understanding writing processes.

\subsection{Automated Classifier Development}
Given the scale of the data, we developed automated classifiers to label it for our codes. This was simple to implement for the majority of codes as all that was required was a string match with the corresponding event name in the data. However, to code for revision behaviors, a more complex algorithm was required (see below). The details of this algorithm, along with paper-related data and analysis can be found in our Github repository\footnote{\url{https://github.com/yixin-cheng/coAuthor}}.

Sentence-level segmentation and analysis allowed us to identify and code the high and low modifications behaviors. By using NLTK sentence tokenizer and identifying ending markers, we collected all sentences in their initial form including their cursor range and ownership (i.e., prompt, user, or api), into a list. Following this, the list was used to track these sentences under the following conditions: \textbf{Sentence Removal}, \textbf{Sentence Merger}, and \textbf{Sentence Revision}. In the case of sentence removal, the corresponding sentence is marked as 'None' in our sentence list. For sentence revisions, the sentence list is updated after every revision. We excluded sentence pairs with a single word difference identified as a misspelling by the \texttt{spellchecker} Python package from being counted as a revision. For sentence mergers, we used TF-IDF to generate word embeddings and calculated the cosine similarity between the updated sentence and the original sentence. The most similar sentence was replaced with the merged version, with others marked as 'None'. The final sentence list's accuracy was corroborated by its exact match with the final story or essay versions from the data, validating our proposed algorithm.

Next, we identified the cursor location of revisions between initial and final sentence lists. The codes \textsc{revise user} and \textsc{revise suggestion}, were identified based on the combination of sentence ownership and updates to the sentence list. To code \textsc{low modification} and \textsc{high modification}, we used Sentence-BERT \cite{Nils2019} to compute sentence cosine similarities between initial and final sentence pairs. We used a threshold value of 0.8 to distinguish high and low modification \cite{Zhang2020}.

To test the reliability of the high/low modification classifiers and the validity of the codes, we randomly sampled 80 events from the data. Two human raters then manually coded these data for the presence/absence of high/low modifications. Next, for both codes we did pairwise comparisons between each set of human coding and the automated classifications. The codes were considered valid and reliable if all pairs of ratings achieved a Cohen's kappa greater than 0.65 and Shaffer's rho less than 0.05. All kappa and rho values for these two codes exceeded these thresholds. The end result of the automated coding process was a dataset with binary values that indicated the presence or absence of a given code for a given event. 

\subsection{Epistemic Network Analysis}
\label{sub:ENA}

To analyze the coded dataset, we used the ENA implementation for R \cite{Marquart}. ENA creates a separate network representation for each unit of analysis where the nodes of the networks are codes and edges between nodes indicate the relative frequency of co-occurrence between those codes in a given unit's data. To construct these networks, we used the following ENA specifications: 

\begin{itemize}
    \item{\textbf{Units of analysis}}: A separate network was created for each combination of session ID and user ID. 
    \item{\textbf{Codes}}: All codes listed in Table \ref{tab:codebook}.
    \item{\textbf{Conversations}}: The data was grouped by session ID, user ID, and sentence for co-occurrence accumulation. That is, codes in all events associated with a given session, user, and sentence could co-occur. 
    \item{\textbf{Window Size}}: An "infinite" stanza window was used to identify the co-occurrence between codes within the data for each unit of analysis. This window begins at the first event in a given conversation and expands to include all events within that conversation. In this way, all coded events within the conversation co-occur, but the result is a more fine-grained accumulation of the co-occurrence structure in the data. 
\end{itemize}

ENA projects the networks into a low dimensional embedding space using singular value decomposition to produce \textit{ENA scores} for each network. This space distinguishes networks in terms of the linear combination of co-occurrence variables that explain the most variance in the data. To help interpret the dimensions of this space, ENA co-registers the network graphs in the embedding space such that the position of the nodes and the connections they define align with the most influential co-occurrences on that dimension. Thus, researchers can visually interpret the dimensions according to the connections at the extremes. For more details, see \cite{Bowman2021}.

\subsection{Regression Analysis}
\label{sub:regression}
To directly address our research questions, we conducted a regression analysis of the ENA results that compared the ownership, genre, and temperature conditions. Our models regressed the ENA scores for each dimension on categorical predictors for ownership (user vs. GAI), prompt type (argumentative vs. creative), and temperature (high vs. low). For each writing session, these values were derived from the associated metadata in the \textit{CoAuthor} dataset. 

Authors may have participated in multiple writing sessions. As a result, the ENA scores were nested within participants. In addition, each participant may have written multiple responses to prompts of the same kind (e.g., userID A118B participated in four argumentative writing session and three creative writing sessions), meaning that participants could also be nested into prompt kinds. To accommodate this nested structure, we used \emph{mixed-effects} regression models \cite{Zuur2009}. To test for the effect of nesting, we calculated the intraclass correlation coefficient (ICC) for the ENA scores grouped within participant and prompt using the \texttt{ICC} package for R. Confidence intervals around the ICC scores suggested that nesting was significant for the participant variable, but not the prompt variable. In turn, we included the participant identifier as a random effect (intercept) in all regression models. 

The outputs of interest were the regression coefficients of the ownership prompt type, and temperature variables, which represent the difference between the mean outcome scores of the two levels of each variable. We conducted the regression modelling using the \texttt{lmer} and \texttt{lmerTest} packages for R \cite{Douglas2015,Alexandra2017}. \texttt{lmerTest} computes significance tests for the regression coefficients using Satterthwaite's method \cite{Satterthwaite1946}. Confidence intervals around the regression coefficients were calculated via bootstrapping using the percentile method \cite{Hyndman1996}. Finally, we calculated the effect size of any significant regression coefficients of interest using Cohen's $d$ \cite{cohen1988}.

\subsection{Network Comparisons}
To interpret the dimensions of the ENA embedding spaces and better visualise differences between subgroups, we examined the corresponding mean ENA networks. That is, for given subgroup in the data, we averaged the edge weights of their associated networks and plotted them in the embedding space. To compare any two subgroups, we computed their network difference graphs by subtracting their corresponding edge weights to show which co-occurrences were more frequent in one group relative to the other.

\section{Results}

\subsection{ENA Embedding Space}
\begin{figure*}[h]
  \centering
  \includegraphics[width=0.6\linewidth]{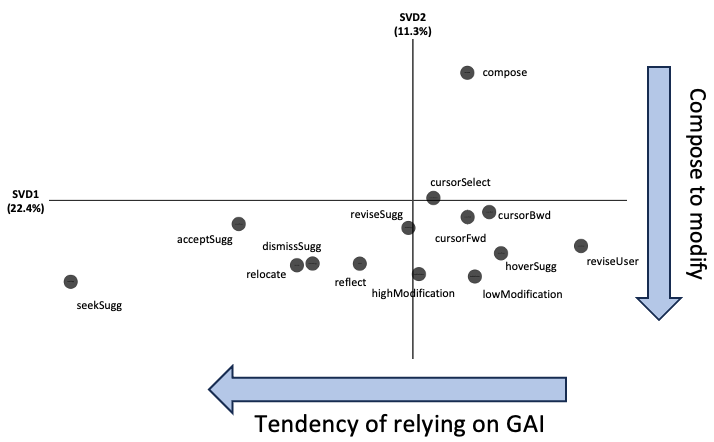}
  \caption{ENA embedding space}
  \label{embdding-space}
\end{figure*}

The resultant ENA embedding space is shown in Figure \ref{embdding-space}. The first dimension primarily distinguishes between authors seeking suggestions from GAI (\textsc{seekSugg}) on the left and revising their own writing (\textsc{reviseUser}) on the right. A shift from left to right reflects the degree to which authors tended to rely on suggestions versus composing and editing their own writing. The second dimension primarily distinguishes between composing (\textsc{compose}) at the top and behaviors related to suggestions and modifications (\textsc{seekSugg}, \textsc{high modification}, and \textsc{low modification}) at the bottom. A higher position along this dimension indicates a greater emphasis on composition, whereas a lower position indicates a greater emphasis on modifications and suggestion use. The dimensions of the space account for 22.4\% and 11.3\% of the total variation in the data, respectively.

\subsection{Regression Results}

\begin{table}
\begin{center}
\begin{tabular}{l c c}
\toprule
& \textbf{X} & \textbf{Y} \\
\midrule
Intercept                   & $-0.03$               & $0.05$                 \\
                            & $(0.04)$              & $(0.03)$               \\
prompt types\_creative              & $0.01$                & $\mathbf{-0.06}^{***}$ \\
                            & $(0.02)$              & $(0.01)$               \\
ownership\_high              & $\mathbf{0.09}^{***}$ & $-0.00$                \\
                            & $(0.02)$              & $(0.02)$               \\
temperature\_high            & $0.02$                & $-0.02$                \\
                            & $(0.02)$              & $(0.01)$               \\
\midrule
AIC                         & $569.90$              & $33.03$                \\
BIC                         & $601.52$              & $64.64$                \\
Log Likelihood              & $-278.95$             & $-10.51$               \\
Num. obs.                   & $1435$                & $1435$                 \\
Num. groups: author\_id     & $60$                  & $60$                   \\
Var: worker\_id (Intercept) & $0.07$                & $0.02$                 \\
Var: Residual               & $0.08$                & $0.05$                 \\
\bottomrule
\multicolumn{3}{l}{\scriptsize{$^{***}p<0.001$; $^{**}p<0.01$; $^{*}p<0.05$}}
\end{tabular}
\caption{Regression models with ENA scores on either dimension as the dependent variable. Standard errors in parentheses.}
\label{table:coefficients}
\end{center}
\end{table}

The results of the regression analysis are shown in Table \ref{table:coefficients}. We report two models that include \textit{author\_id} as a random effect, \textit{prompt types}, \textit{ownership}, and \textit{temperature} as fixed effects, and ENA scores on either the first or second dimension of the ENA space as the outcome. Testing for interactions between the fixed effects did not yield significant results, so we report only the main effects here. 

On the first dimension (X), only the ownership variable was significant. Authors with more ownership over their final written product (user ownership) were significantly higher on the dimension than those with less ownership (GAI ownership): $t = 4.35$, $p < 0.001$, $d = 0.24$, 95\% CI $[0.05,0.13]$. Given the interpretation of the first dimension above, this result suggests that user ownership authors tended to focus on composing and revising their own writing significantly more than GAI ownership authors. 

On the second dimension (Y), only the prompt condition variable was significant. When the authors responded to creative writing prompts, they were significantly lower on the dimension compared to when they responded to argumentative prompts: $t = -4.71$, $p < 0.001$, $d = -0.22$, 95\% CI $[-0.09,-0.04]$. This result suggests that when the authors responded to creative writing prompts, they tended to focus on editing their writing and using GAI suggestions significantly more than when they responded to argumentative writing prompts. 

The temperature variable was not statistically significant on either dimension suggesting that we have insufficient empirical evidence to falsify the null hypothesis of no difference between the high and low temperature conditions. 

\subsection{User vs. GAI ownership}
\label{sub: ownership}

\begin{figure*}[h]
  \centering
  \includegraphics[width=\linewidth]{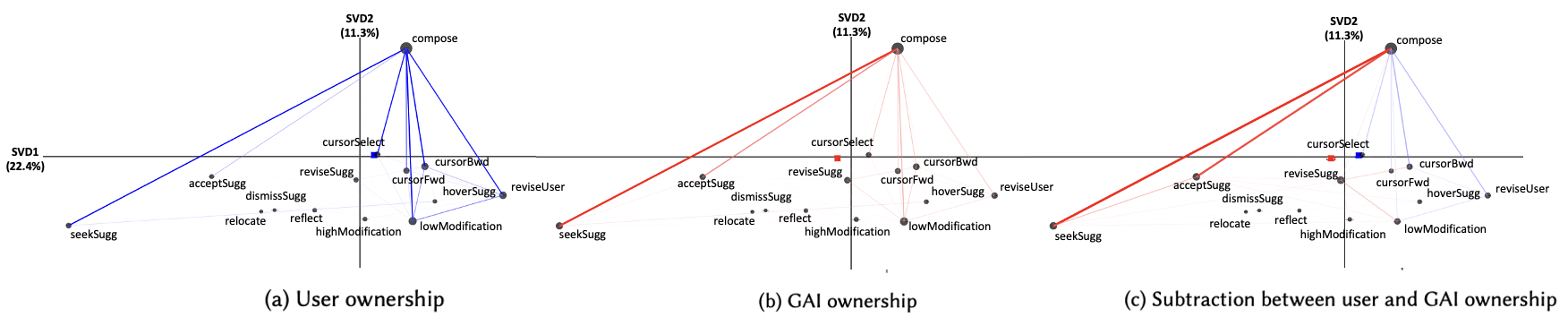}
  \caption{Ownership}
  \label{fig:ownership}
\end{figure*}

Figure \ref{fig:ownership} shows the network comparison for the \emph{ownership} variable; user ownership on the left (a) and GAI ownership in the middle (b). Both networks include a large number of connections to the \textsc{compose} node suggesting that authors tended to link their composition process to a variety of other behaviors. For example, both networks show relatively strong connections between \textsc{compose} and \textsc{seekSugg}, \textsc{acceptSugg}, \textsc{reviseUser}, and cursor movements. The network differences are shown in plot (c). Here, red edges indicate more frequent co-occurrences in the GAI ownership group; blue edges indicate more frequent co-occurrences in the user ownership group. The graph indicates that the GAI ownership authors made stronger connections between \textsc{compose} and \textsc{seekSugg}, \textsc{compose} and \textsc{acceptSugg}, and \textsc{reviseSugg} and \textsc{lowModification}. User ownership authors, on the other hand, made stronger connections between \textsc{compose} and cursor movements and \textsc{compose} and \textsc{reviseUser}. The centroids of the networks---that is average ENA scores (blue and red squares)---corroborate the regression analysis, with the GAI ownership authors being further to the left on the first dimension than the user ownership authors, on average.

\subsection{Creative vs. Argumentative}
\label{sub:prompt types}

\begin{figure*}[h]
  \centering
  \includegraphics[width=\linewidth]{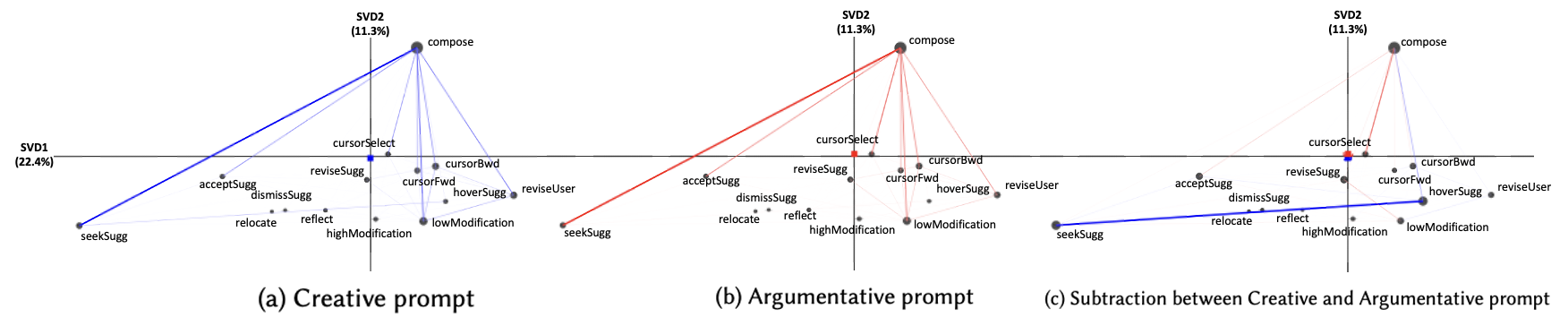}
  \caption{Prompt type}
  \label{fig:prompt types}
\end{figure*}

Figure \ref{fig:prompt types} shows the network comparisons for the \textit{prompt types} variable; the mean network for responses to creative writing prompts is on the left (a), and the mean network for responses to argumentative writing prompts is in the middle (b). As before, in both networks, connections to the \textsc{Compose} node feature prominently.
The subtracted network is shown in subplot (c). Here, blue edges represent connections that occurred more frequently in the creative condition and red edges represent connections that occurred more frequently in the argumentative condition. The graph indicates that when responding to creative writing prompts, authors tended to explore AI generated suggestions more, having stronger connections among \textsc{seekSugg}, \textsc{hoverSugg}, and \textsc{compose}. When responding to argumentative prompts, authors tended to focus more on composing their own writing and revising AI-generated suggestions---as indicated by stronger connections between \textsc{compose} and \textsc{cursorSelect}, \textsc{compose} and \textsc{acceptSugg}, and \textsc{reviseSugg} and \textsc{lowModification}. The centroids of the two networks corroborate the regression analysis, with the argumentative condition being higher on the second dimension than the creative condition, on average.

\subsection{High vs. Low Temperature}
\label{sub:temp}
\begin{figure*}[h]
  \centering
  \includegraphics[width=\linewidth]{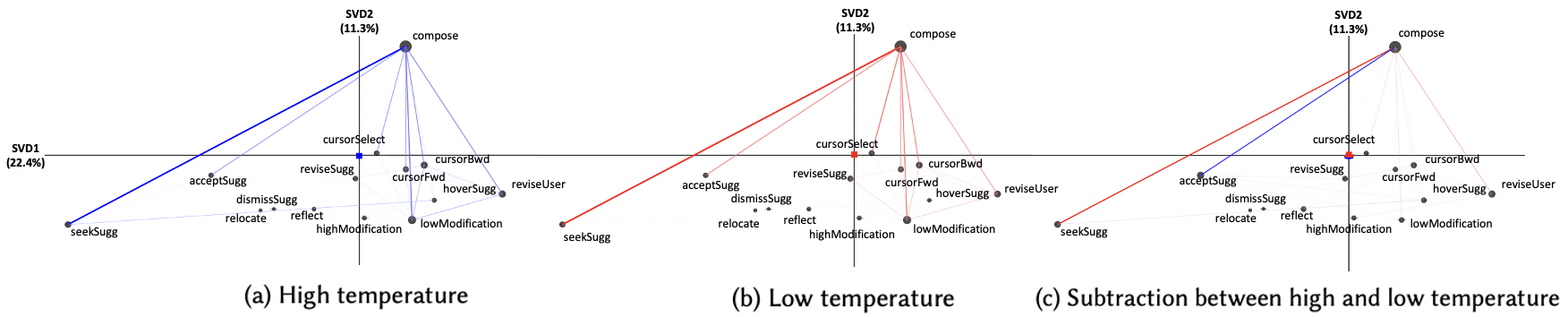}
  \caption{Temperature}
  \label{fig:temp}
\end{figure*}

Figure \ref{fig:temp} contrasts for the networks for the high (a) and low (b) temperature conditions. The network subtraction (c) indicates that the authors in the high temperature condition made stronger connections between \textsc{compose} and \textsc{seekSugg}, \textsc{compose} and \textsc{reviseUser}, and \textsc{compose} and cursor movements. The authors in the high temperature condition made stronger connections between \textsc{compose} and \textsc{acceptSugg}. The centroids of the two networks overlap on both dimensions, corroborating the regression results of no significant difference between the two conditions. 

\section{Discussion and Conclusions}
\label{sec:discussion}
In this study, we sought to demonstrate an assessment method for human-AI collaborative writing. Framed by ECD, we used elements of knowledge-telling, knowledge transformation, and cognitive presence to identify claims for our student model; we used data collected from the \textit{CoAuthor} writing tool as a proxy for our task model; and we used ENA to evidence claims about the student model using data from the task model. More specifically, we compared the co-writing behaviors of users across three conditions: high vs. low ownership; creative vs. argumentative prompt types; and high vs. low temperature. We found statistically significant differences between the process of authors in the high/low ownership conditions and the creative/argumentative conditions. Specifically, authors with GAI ownership over their final product tended to rely more on GAI suggestions, while those with user ownership tended to focus more on composing and revising their own writing. When responding to creative writing prompts, authors tended to explore GAI suggestions more, while those responding to argumentative prompts tended focus on composing their own writing and making small revisions to GAI writing. 

In terms of more specific assessment claims, the results support \textbf{Hypothesis 1}---that frequent users of GAI outputs would tend more towards knowledge telling and triggering events, as evidenced by stronger connections between \textsc{compose} and \textsc{seekSugg}, as well as \textsc{compose} and \textsc{acceptSugg}. \textbf{Hypothesis 2}---that less frequent users of GAI would tend toward knowledge transformation, exploration, integration, and resolution was not supported. Authors with more ownership over their written product sought and accepted some GAI suggestions but did not tend to revise them. Instead their revisions tended to be made on their own text with only slight modifications as evidenced by stronger connections among \textsc{compose}, \textsc{reviseUser}, and \textsc{lowModification}.

The results were mixed for \textbf{Hypothesis 3}---that authors responding to creative writing prompts would tend toward knowledge telling, triggering events, and exploration. On the one hand, they did focus more on triggering events and exploration, as evidenced by stronger connections between \textsc{seekSugg} and \textsc{hoverSugg}, as well as \textsc{compose} and \textsc{hoverSugg}. On the other hand, evidence of a greater focus on knowledge telling is less clear. Authors that responded to argumentative prompts had stronger connections between \textsc{acceptSugg} and \textsc{compose}. But they also had stronger connections between \textsc{reviseSugg} and \textsc{lowModification}, indicating at least some level of transformation that was less prevalent for those responding to creative writing prompts. The results were similarly mixed for \textbf{Hypothesis 4}---that authors responding to argumentative writing prompts would be characterized by knowledge transformation, integration, and resolution. There is some evidence for this given stronger connections between \textsc{reviseSugg} and \textsc{lowModification}, but the lack of connections to \textsc{highModificaiton} and \textsc{reflect} limit this interpretation.

\textbf{Hypothesis 5} and \textbf{Hypothesis 6}---that authors interacting with lower temperature GAI would tend more toward knowledge transformation, integration, and resolution, while authors interacting with higher temperature GAI would tend more toward knowledge telling and exploration---were not supported. While authors in the high temperature condition did have stronger connections to \textsc{acceptSugg} and authors in low temperature condition had stronger connections to \textsc{seekSugg}, their overall co-occurrence patterns were highly similar. 

Our study has several limitations. First, as with any study, our results are limited by the data at hand. In this case, our data comes from self-selected participants who engaged with the \textit{CoAuthor} platform. This potentially introduces a selection bias, as these individuals represent a subset with specific characteristics or preferences, which might not be fully representative of the wider population. 

Second, the tasks and environment we used are not necessarily representative of the broader selection of GAI tools for writing. \textit{CoAuthor} constrains user interactions with GAI by only allowing them to seek suggestions that continue the current text. Widely used tools, however, afford less restricted interactions where users can phrase questions to the GAI any way they want, as well as include "steering" instructions that tell the GAI how to respond in general. Moreover, the available \textit{CoAuthor} dataset contains interactions with the now outdated GPT-3. It is possible that interactions with more contemporary versions of GAI might yield different results. 

Third, our coding scheme, and thus our proposed student model, only focused on a narrow subset of cognitive aspects of writing related to observable behaviors in the data. Our analysis did not consider other potentially important features such as the semantic content of the writing and the metacognitive strategies being used. Future studies will explore these as possible assessment targets.

Finally, our proposed assessment method is plausible but its inherent complexity could restrict its scalability and accessibility. We aim to address these limitations by developing an adaptable and flexible API that includes customizable parameters, such as event names, to meet diverse task requirements such as post-event or real-time analysis in human-AI writing environments. This solution seeks to bridge the gap between the current proof of concept and its practical, large-scale application.

Despite these limitations, our work provides a proof of concept for the evidence-centered assessment of writing composed with the help of GAI. Our methodology posits specific features of human-AI collaborative writing to target; adopts an existing task model to produce assessment data; and leverages process models to relate these data to differences between the writing processes of participants. We hope that this work will continue to expand such that we have assessments not just for writing, but a variety of meaningful interactions between learners and AI. Such assessments will be crucial to preparing learners for an age in which effective human-AI collaboration is an essential skill.

\section*{Acknowledgments}
This research was in part supported by the Australian Research Council (DP220101209, DP240100069) and Jacobs Foundation (CELLA 2 CERES).

\section*{Citation}
Yixin Cheng, Kayley Lyons, Guanliang Chen, Dragan Gašević, and Zachari Swiecki. 2024. Evidence-centered Assessment for Writing with Generative AI. In The 14th Learning Analytics and Knowledge Conference (LAK ’24), March 18–22, 2024, Kyoto, Japan. ACM, New York, NY, USA, 16 pages. https://doi.org/10.1145/3636555.3636866

\bibliographystyle{unsrt}  
\bibliography{references}

\end{document}